\documentstyle[12pt]{article}

\begin{document}
\setcounter{page}{1}
\pagestyle{plain} \vspace{1cm}
\begin{center}
\Large{\bf  Vacuum expectation value profiles of the bulk scalar field in the generalized Randall-Sundrum model}\\
\small \vspace{1cm} { {\bf A.
Tofighi$^a$}},\footnote{A.Tofighi@umz.ac.ir}\quad {\bf M.
Moazzen$^b$}, { {\bf A. Farokhtabar$^a$}} \\
\vspace{0.5cm} {\it $^a$Department of Physics, Faculty of Basic Sciences,\\
University of Mazandaran,
P. O. Box 47416-95447, Babolsar, Iran.}\\
\vspace{0.5cm} {\it $^b$Young Researchers and Elite Club, Bojnourd Branch,\\
 Islamic Azad University, Bojnurd, Iran.}\\
\end{center}
\vspace{1.5cm}
\begin{abstract}
In the generalized Randall-Sundrum warped braneworld model the
cosmological constant induced on the visible brane can be positive
or negative. In this paper we investigate profiles of vacuum
expectation value of the bulk scalar field under general Dirichlet
and Neumann boundary conditions in the generalized warped braneworld
model. We show that the VEV profiles generally depends on the value
of the brane cosmological constant. We find that the VEV profiles of
the bulk scalar field for a visible brane with negative cosmological
constant and positive tension is quite distinct from those of
Randall-Sundrum model. In addition we show that the VEV profiles for
a visible brane with large positive cosmological constant are also
different from those of the Randall-Sundrum
model.\\
We also verify that Goldberger and Wise mechanism can work under non-zero Dirichlet boundary conditions in the generalized Randall-Sundrum model.
\end{abstract}
\vspace{1.5cm}

\newpage
\section{Introduction}
Extra dimension is an important subject in the realm of theoretical
physics that provides many creative ways to solve some problem in
physics such as hierarchy problem. Unlike the model of Arkani-Hamed,
et. al.$[1]$, Randall and Sundrum (RS) proposed an alternative
scenario $[2]$ to solve the hierarchy problem that does not require
large extra dimensions. They assumed an exponential function of the
compactification radius called a warp factor in a $5-$dimensional
anti-de sitter space-time compactified on a $S^1/Z_2$ orbifold. Two
3-branes are located at the orbifold fixed point $\phi=\pi$ (visible
brane) and $\phi=0$ (hidden brane). Due to the non-factorizable
geometry of metric all fundamental scalar masses
are subject to an exponential suppression on the visible brane.\\
In the original RS model the cosmological constant induced on the
visible brane is zero and the brane tension is negative. This model
has been generalized such that the cosmological constant on the
brane as well as brane tension can be positive or negative $[3]$. In
the RS model the size of extra dimension, $r_{c}$, is not determined
by the dynamic of the model. For this scenario to be relevant, it is
necessary to find a mechanism for generating a potential to
stabilize the value of $r_{c}$. This mechanism was proposed by
Goldberger and Wise (GW) $[4]$ so that the dynamic of a five
dimensional bulk scalar field in such model could stabilize the size
of extra dimension. In the GW mechanism the potential for the radion
that sets the size of the fifth dimension is generated by a bulk
scalar with quartic interactions localized on the two $3-$branes.
The minimum of this potential yields a compactification scale that
solves the hierarchy problem without fine tuning of parameters. The
back reaction of the bulk scalar field was neglected in the original
GW
mechanism but it was studied in $[5]$ later. Some studies about this mechanism can be found in $[6-8]$.\\
Recently, GW mechanism of radius stabilization in the brane-world
model with non zero brane cosmological constant has been considered
$[9]$. It was shown that for a generalized RS model, the modulus
stabilization condition explicitly depends on the brane cosmological
constant. Furthermore Haba, et. al. $[10]$ have analyzed profiles of
vacuum expectation value (VEV) of the bulk scalar field under the
general boundary conditions (BCs) on the RS warped compactification.
They have investigated GW mechanism in several setups with the
general BCs of the bulk scalar field. Also they showed that
$SU(2)_R$ triplet Higgs in the bulk left-right symmetric model with
custodial symmetry can be identified the Goldberger-Wise scalar.\\
The motivation for the present study is to study the vev profile of
the scalar field, including a brane cosmological constant and with
different combinations of Dirichlet and Neumann boundary conditions
at the two branes.  We also want to know if this scalar can
 stabilize the
size of the warped extra-dimension? We note that he present
accelerating phase of the universe is due to the presence of small
positive cosmological constant with a tiny value of $10^{-124}$ in
Planck unit. Hence it is desirable to consider the effect of
non-zero cosmological constant on the brane.\\
 This paper is organized as follows. In section $2$, we briefly
summarize the braneworld model with non zero brane cosmological
constant. In section $3$ we study the behavior of the bulk scalar
field under four BCs on the general RS warped model in the case
without brane localized potential and in the next section we analyze
profiles of VEV of the bulk scalar field in the case with brane
localized scalar potential. Also we investigate the GW mechanism
under the BCs in the generalized braneworld scenario. Finally in
section $5$ we conclude with the summary of our
results.\\
\section{The generalized Randall-Sundrum model}
One of the many different possibilities that can be explored to
solve the hierarchy problem is Randall-Sundrum (RS) model $[2]$. The
RS model proposes that spacetime is described by a $5D$ Anti-de
Sitter (AdS) metric. Some good reviews on this subject can be found
in, e.g., Refs. $[11-13]$. In the Randall-Sundrum model the visible
brane tension is negative and the cosmological constant on the
visible brane assumed to be zero. It has been shown in $[3]$ that
one can indeed generalize the model with nonzero cosmological
constant on the brane and still can have Planck to TeV scale warping
from the resulting warp factor. In this section we want to study the
warped braneworld model with non zero brane cosmological constant
briefly. Other studies for this model have been reported in
$[14-16]$.\\ The action of a bulk scalar field, $\Phi$, on the warp
braneworld model that suggested by Randall and Sundrum is
\begin{equation}
S=\int d^{5}x
\sqrt{-G}[-G^{MN}(\partial_{M}\Phi^{\dag})(\partial_{N}\Phi)-V].
\end{equation}
Where $x^{M}=(x^{\mu},y)$, $\mu=0,1,2,3$. The general form of the
warped metric for a five dimensional space-time is given by
\begin{equation}
G_{MN}dx^{M}dx^{N}=e^{-2{A(y)}}g_{\mu\nu}dx^{\mu}dx^{\nu}+dy^{2},
\end{equation}
where $g_{\mu\nu}$ stands for four dimensional curved brane. The
brane tension $k$ is defined by
\begin{equation}
k\equiv\pm\sqrt{\frac{-\Lambda}{6M_{5}^{3}}}=\frac{V_{UV}}{6M_{5}^{3}}=-\frac{V_{IR}}{6M_{5}^{3}},
\end{equation}
where $M_{5}$ is the Planck mass in five-dimensions. We assume that
the scalar field $\Phi$ is a function of the extra dimension $y$
only that can be defined as $[10]$
\begin{equation}
\Phi=\frac{\Phi_{R}+i\Phi_{I}}{\sqrt{2}},
\end{equation}
and for simplicity, we take the bulk potential as
\begin{equation}
V=m^{2}|\Phi|^{2}=\frac{m^{2}}{2}(\Phi_{R}^{2}+\Phi_{I}^{2}).
\end{equation}
 A scalar mass on the visible
brane gets warped through the warp factor $e^{-A(kr\pi)}=10^{-n}$
where the warp factor index, $'n'$, is set to $16$ to achieve the
desired warping from Planck to TeV scale. The magnitude of the
induced cosmological constant on the brane in the generalized RS
model is non-vanishing and is given by $10^{-\aleph}$ in planck
unit. For the negative value of the cosmological constant,
$\Omega<0$, on the visible brane which leads to AdS-Schwarzschild
space-times the solution of the warp factor can be written as $[9]$
\begin{equation}
e^{-A(y)}=\omega \cosh(\ln\frac{\omega}{c_{1}}+ky),
\end{equation}
where $\omega^{2}\equiv\frac{-\Omega}{3k^{2}}\geq0$ and
$c_{1}=1+\sqrt{1-\omega^{2}}$. It was shown that real solution for
the warp factor exists if and only if $\omega^{2}\leq 10^{-2n}$.
This leads to an upper bound for the magnitude of the cosmological
constant as $\aleph=2n$. When $\aleph=\aleph_{min} = 2n$, the
visible brane tension is zero. For AdS brane, there are degenerate
solutions of $kr\pi$ whose values will depend on $\omega^{2}$ and
$'n'$. The brane tension is negative for some of these solutions and
is positive for others. Next, for AdS brane we want to investigate
the solution of the equation of motion which is obtained from the
action given by $eq.(1)$. As it has been discussed in $[3]$, to
resolve the gauge hierarchy problem without introducing any
intermediate scale, the brane cosmological constant should be tuned
to a very small value therefore in this case the warp factor in
$eq.(6)$ can be given by $[9]$
\begin{equation}
e^{-4A(y)}=e^{-4ky}+\omega^{2}e^{-2ky}.
\end{equation}
We use background field method to study behaviors of the bulk scalar
field of the generalized Randall-Sundrum model. In this method one
 separates the field into classical and quantum fluctuation parts.
The configuration of the classical field obeys an equation of
motion $[10]$.\\
With the above warp factor equation of motion for the classical
field can be written down
\begin{equation}
\partial_{y}^{2}\Phi_{X}-(4k-2k\omega^{2}e^{2ky})\partial_{y}\Phi_{X}-m^{2}\Phi_{X}=0,
\end{equation}
where $X$ stands for $R$ and $I$. We solve the above equation and
obtain the solution for the scalar field  as
\begin{equation}
\Phi_{X}(z)=Az^{\nu+2}+Bz^{-(\nu-2)}+\alpha Az^{\nu+4}+\beta Bz^{-(\nu-4)},
\end{equation}
where
\begin{equation}
\alpha=\frac{-\omega^{2}}{2}(\frac{\nu+2}{\nu+1}),\quad\quad\quad
\beta=\frac{-\omega^{2}}{2}(\frac{2-\nu}{1-\nu}),
\end{equation}
and $\nu=\sqrt{4+\frac{m^{2}}{k^{2}}}$ and $z=e^{ky}$. In the above
solution $A$ and $B$
 are arbitrary constant which evaluated by using the appropriate boundary conditions at the locations of the
 brane.\\
 For the positive induced brane cosmological constant $\Omega>0$ which corresponds to
dS-Schwarzschild space-time the warp factor turns out to be
\begin{equation}
e^{-A(y)}=\omega \sinh(\ln\frac{c_{2}}{\omega}-ky),
\end{equation}
where $\omega^{2}\equiv\frac{\Omega}{3k^{2}}\geq0$ and
$c_{2}=1+\sqrt{1+\omega^{2}}$. In this case there is no bound on the
value of $\omega^{2}$, and the positive cosmological constant can be
of arbitrary magnitude. For dS brane, the brane tension is negative
for the entire range of values of the positive cosmological
constant. Also for a small $\omega$ the warp factor can be written
down
\begin{equation}
e^{-4A(y)}=e^{-4ky}-\omega^{2}e^{-2ky}.
\end{equation}
By using the above warp factor the solution of the equation of
motion for the dS brane is
\begin{equation}
\Phi_{X}(z)=Az^{\nu+2}+Bz^{-(\nu-2)}-\alpha Az^{\nu+4}-\beta Bz^{-(\nu-4)}.
\end{equation}
It is obvious that by taking $\alpha\rightarrow -\alpha$ and
$\beta\rightarrow -\beta$ we can obtain the above solution from AdS
one. Notice that from now on we will use $\omega^{2}$ to represent
induced brane cosmological constant. With this generalized RS warped
model we now investigate the profile of the bulk scalar field under
boundary conditions in the next section.

\section{ VEV profiles in the absence of brane localized potential}
In this section, we study the VEV profiles of the bulk scalar field
in a case without brane localized scalar potential in the
generalized Randall-Sundrum model. By utilizing the $eq.(2)$, the
action can be rewritten as
\begin{equation}
S=\int d^{4}x\int_{0}^{L} dy
e^{-4A(y)}[-e^{2A(y)}|\partial_{\mu}\Phi|^{2}-|\partial_{y}\Phi|^{2}-V].
\end{equation}
The above action was defined on a line segment as $0\leq y \leq L$.
If we write the bulk scalar as $eq.(4)$ then the variation of the
action is given by
\begin{equation}
\delta S=\int d^{4}x\int_{0}^{L} dy e^{-4A(y)}[\delta \Phi_{X}(\rho
\Phi_{X}-\frac{\partial V}{\partial \Phi_{X}})+ \delta (y)\delta
\Phi_{X}(+ \partial_{y}\Phi_{X})+\delta (y-L)\delta \Phi_{X}(-
\partial_{y}\Phi_{X})].
\end{equation}
Where $\rho =e^{2A(y)}\Box + e^{4A(y)}
\partial_{y}e^{-4A(y)}\partial_{y} $. The VEV profile is obtained by
the action principal $\delta S=0$, that is
\begin{equation}
(\rho \Phi_{X}-\frac{\partial V}{\partial \Phi_{X}})=0.
\end{equation}
Notice that the above equation of motion has been solved in the
pervious section for dS and AdS brane in $eq.(9)$ and $eq.(13)$
respectively. The boundary conditions at $y=0$ and $L$ reads either
Dirichlet
\begin{equation}
\delta \Phi_{X}|_{y=0,L}=0,
\end{equation}
or Neumann
\begin{equation}
\pm \partial_{y} \Phi_{X}|_{y=0,L}=0.
\end{equation}
Where plus and minus signs are for $y=0$ and $y=L$ respectively. As
mentioned in $[10]$ we can have four choices of combination of
Dirichlet and Neumann boundary conditions showed by $(D,D)$,
$(D,N)$, $(N,D)$ and $(N,N)$ at $y=0$ and $L$. Here by using the
solution of the equation of motion that given by $eq.(9)$ and
$eq.(13)$ we want to verify the profile of the bulk scalar field
under the four boundary conditions in the generalized Randall-
Sundrum model with non-zero brane cosmological constant.

\subsection{$(D,D)$ case}
We discuss a case in which both boundary conditions on the $y = 0$
and $y = L$ branes are the Dirichlet type boundary conditions. The
most general form of the Dirichlet BC is $\delta \Phi|_{z=\xi}=0$
and
\begin{equation}
\upsilon(1)=\upsilon_{1},\quad\quad\quad
\upsilon(z_{L})=\upsilon_{2},
\end{equation}
where $\xi$ is taken as $1$ and $z_{L}=e^{kL}$. For AdS brane these
boundary conditions can be rewritten by $eq.(9)$ as
\begin{equation}
A(1+\alpha)+B(1+\beta)=\upsilon_{1}\quad\quad\quad
Az_{L}^{\nu+2}(1+\alpha z_{L}^{2})+Bz_{L}^{(-\nu+2)}(1+\beta
z_{L}^{2})=\upsilon_{2}.
\end{equation}
The above equations lead to
\begin{equation}
A=-\frac{\upsilon_{1}z_{L}^{(-\nu+2)}(1+\beta
z_{L}^{2})-\upsilon_{2}(1+\beta)}{z_{L}^{(\nu+2)}(1+\alpha
z_{L}^{2})(1+\beta)-z_{L}^{(-\nu+2)}(1+\beta z_{L}^{2})(1+\alpha)},
\end{equation}
and
\begin{equation}
B=\frac{\upsilon_{1}z_{L}^{(\nu+2)}(1+\alpha
z_{L}^{2})-\upsilon_{2}(1+\alpha)}{z_{L}^{(\nu+2)}(1+\alpha
z_{L}^{2})(1+\beta)-z_{L}^{(-\nu+2)}(1+\beta z_{L}^{2})(1+\alpha)}.
\end{equation}
Therefore, under $(D,D)$ boundary conditions, in the generalized RS
model with negative brane cosmological constant the VEV profile is\\

\quad\quad\quad\quad\quad\quad$$\upsilon(z)=-\frac{\upsilon_{1}z_{L}^{(-\nu+2)}(1+\beta
z_{L}^{2})-\upsilon_{2}(1+\beta)}{z_{L}^{(\nu+2)}(1+\alpha
 z_{L}^{2})(1+\beta)-z_{L}^{(-\nu+2)}(1+\beta
 z_{L}^{2})(1+\alpha)}z^{\nu+2}$$\\
$$+\frac{\upsilon_{1}z_{L}^{(\nu+2)}(1+\alpha
z_{L}^{2})-\upsilon_{2}(1+\alpha)}{z_{L}^{(\nu+2)}(1+\alpha
 z_{L}^{2})(1+\beta)-z_{L}^{(-\nu+2)}(1+\beta
 z_{L}^{2})(1+\alpha)}z^{-\nu+2}$$\\
$$-\alpha(\frac{\upsilon_{1}z_{L}^{(-\nu+2)}(1+\beta
z_{L}^{2})-\upsilon_{2}(1+\beta)}{z_{L}^{(\nu+2)}(1+\alpha
 z_{L}^{2})(1+\beta)-z_{L}^{(-\nu+2)}(1+\beta
 z_{L}^{2})(1+\alpha)})z^{\nu+4}$$
\begin{equation}
+\beta(\frac{\upsilon_{1}z_{L}^{(\nu+2)}(1+\alpha
z_{L}^{2})-\upsilon_{2}(1+\alpha)}{z_{L}^{(\nu+2)}(1+\alpha
z_{L}^{2})(1+\beta)-z_{L}^{(-\nu+2)}(1+\beta
z_{L}^{2})(1+\alpha)})z^{-\nu+4},
\end{equation}
notice that for $\omega=0$ (i.e., the RS case) which corresponds to
$\alpha=0$ and $\beta =0$ above equations lead to the results that
proposed by Haba $[10]$ for $(D,D)$ BCs.

\begin{figure}[htp]
 \begin{center}\includegraphics{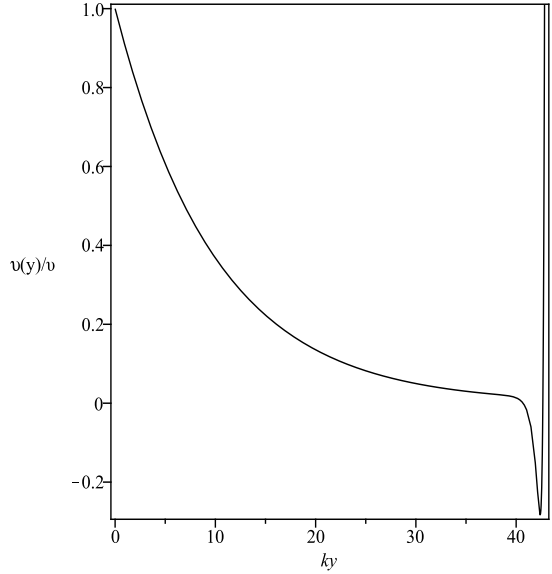}\vspace{6cm}
 \end{center}
  \caption{\small {The VEV profile under $(D,D)$ BCs as a function of extra
dimension for $\aleph=34$ in the absence of brane potential with
$\Omega < 0$ and a
positive tension visible brane.
We take $\upsilon\equiv\upsilon_{0,1,2,L}\equiv M^{\frac{3}{2}}_{pl}$, $k\equiv M_{pl}$, $\nu=2.1$, $M_{pl}=2.4\times10^{18}GeV$
.}}
 \end{figure}

 In figure $1$ the VEV
profile  for $\aleph=34$ is shown for the $AdS$ brane. In this figure
we have assumed
the brane tension of visible brane to be positive (see $eq.(66)$ for detail).
 we find
that the pattern of localization is different from that of the $RS$
case. It is seen that the position of the IR brane shifts to the
larger value of $z_{L}$ so the drastic change of the VEV profile
occurs later than RS case. For this figure  we have taken $\upsilon\equiv\upsilon_{0,1,2,L}\equiv M^{\frac{3}{2}}_{pl}$, $k\equiv M_{pl}$, $\nu=2.1$, $M_{pl}=2.4\times10^{18}GeV$ .\\
For dS brane ,$\alpha\rightarrow -\alpha$ and $\beta\rightarrow
-\beta$ therefore $A$ and $B$ can be written as
\begin{equation}
A=-\frac{\upsilon_{1}z_{L}^{(-\nu+2)}(1-\beta
z_{L}^{2})-\upsilon_{2}(1-\beta)}{z_{L}^{(\nu+2)}(1-\alpha
z_{L}^{2})(1-\beta)-z_{L}^{(-\nu+2)}(1-\beta z_{L}^{2})(1-\alpha)},
\end{equation}
and
\begin{equation}
B=\frac{\upsilon_{1}z_{L}^{(\nu+2)}(1-\alpha
z_{L}^{2})-\upsilon_{2}(1-\alpha)}{z_{L}^{(\nu+2)}(1-\alpha
z_{L}^{2})(1-\beta)-z_{L}^{(-\nu+2)}(1-\beta z_{L}^{2})(1-\alpha)}.
\end{equation}
By utilizing the above form of $A$ and $B$ in $eq.(13)$ the VEV
profile of bulk scalar field is obtained for dS brane. We
investigate this VEV profile for large and small values of $\omega$.
We find that for the small values of positive brane cosmological
constant the VEV profile has tiny difference with respect to the RS
case.\\
\begin{figure}[htp]
 \begin{center}\includegraphics{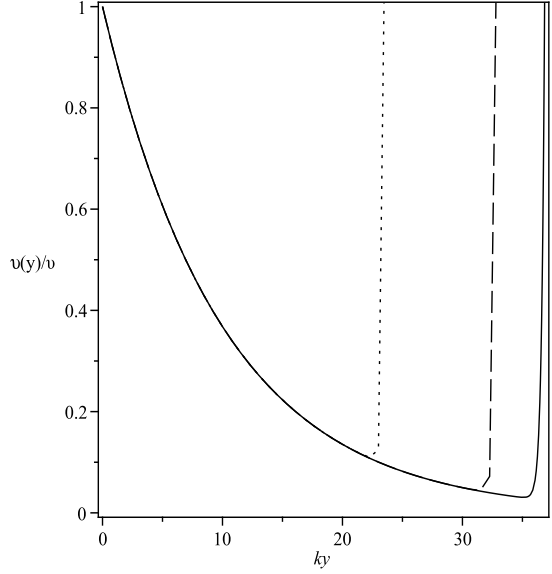}\vspace{6cm}
 \end{center}
  \caption{\small {The dependence of VEV profile under $(D,D)$ BCs
for large positive brane cosmological constant with $\aleph=20$
(dotted line), $\aleph=28$ (dashed line) and $\aleph=\infty$ (solid
line i.e., RS case)
. We take the same values for other parameter as
ones in Figure $1$
.}}
 \end{figure}
 In figure.$2$ we show the results for the large positive
brane cosmological constant, namely $\aleph=20$ (dotted line) and
$\aleph=28$ (dashed line). For the sake of comparison we also
include  $\aleph=\infty$ (solid line i.e., RS case). It is observed
that the profile localizes near the IR brane. But the location of
the IR brane is different in comparison to the RS case (see
$eq.(69)$ of the appendix.\\
In $(D,N)$ case, the boundary conditions can be given by
\begin{equation}
\upsilon(1)=\upsilon_{1},\quad\quad\quad\partial_{z}
\upsilon(z)|_{z=z_{L}}=0.
\end{equation}
Then, for AdS brane they are written down by
\begin{equation}
A(1+\alpha)+B(1+\beta)=\upsilon_{1},
\end{equation}
and
\begin{equation}
Az_{L}^{\nu+2}(v+2+\alpha(\nu+4)
z_{L}^{2})-Bz_{L}^{-(\nu-2)}(\nu-2+\beta(\nu-4) z_{L}^{2})=0.
\end{equation}
These lead to
\begin{equation}
A=\frac{\upsilon_{1}z_{L}^{(-\nu+2)}[(\nu-2)+\beta
z_{L}^{2}(\nu-4)]}{z_{L}^{(-\nu+2)}(1+\alpha)[(\nu-2)+\beta
z_{L}^{2}(\nu-4)]+z_{L}^{(\nu+2)}(1+\beta)[(\nu+2)+\alpha
z_{L}^{2}(\nu+4)]},
\end{equation}
and
\begin{equation}
B=\frac{\upsilon_{1}z_{L}^{(\nu+2)}[(\nu+2)+\alpha
z_{L}^{2}(\nu+4)]}{z_{L}^{(-\nu+2)}(1+\alpha)[(\nu-2)+\beta
z_{L}^{2}(\nu-4)]+z_{L}^{(\nu+2)}(1+\beta)[(\nu+2)+\alpha
z_{L}^{2}(\nu+4)]}.
\end{equation}
By substituting the above form of $A$ and $B$ in $eq.(9)$ the VEV
profile in $(D,N)$ case of boundary conditions is obtained.\\
\subsection{$(N,D)$ case}
 Next, we consider the $(N,D)$ case. These boundary conditions can
 be described by
 \begin{equation}
\partial_{z}
\upsilon(z)|_{z=1}=0 ,\quad\quad\quad \upsilon(z_{L})=\upsilon_{2}.
\end{equation}
For negative brane cosmological constant, they are
\begin{equation}
A(\nu+2+\alpha (\nu+4))-B(\nu-2+\beta (\nu-4))=0,
\end{equation}
and
\begin{equation}
Az_{L}^{\nu+2}(1+\alpha z_{L}^{2})+Bz_{L}^{(-\nu+2)}(1+\beta
z_{L}^{2})=\upsilon_{2}.
\end{equation}
 These lead to
\begin{equation}
A=\frac{\upsilon_{2}[(\nu-2)+\beta(\nu-4)]}{z_{L}^{(-\nu+2)}(1+\beta
z_{L}^{2})[(\nu+2)+\alpha(\nu+4)]+z_{L}^{(\nu+2)}(1+\alpha
z_{L}^{2})[(\nu-2)+\beta (\nu-4)]},
\end{equation}
and
\begin{equation}
B=\frac{\upsilon_{2}[(\nu+2)+\alpha(\nu+4)]}{z_{L}^{(-\nu+2)}(1+\beta
z_{L}^{2})[(\nu+2)+\alpha(\nu+4)]+z_{L}^{(\nu+2)}(1+\alpha
z_{L}^{2})[(\nu-2)+\beta (\nu-4)]}.
\end{equation}
Like pervious subsection we substitute the above form of $A$ and $B$
in $eq.(9)$ to obtain the VEV profile in $(N,D)$ case of boundary
conditions.\\
\subsection{$(N,N)$ case}
Finally, we investigate the $(N,N)$ case. These boundary conditions
 are
\begin{equation}
\partial_{z}
\upsilon(z)|_{z=1}=0 ,\quad\quad\quad \partial_{z}
\upsilon(z)|_{z=z_{L}}=0.
\end{equation}
Then, for AdS brane they are written down by
\begin{equation}
A(\nu+2+\alpha(\nu+4))-B(\nu-2+\beta(\nu-4))=0,
\end{equation}
and
\begin{equation}
Az_{L}^{\nu+2}(\nu+2+\alpha(\nu+4)
z_{L}^{2})-Bz_{L}^{-(\nu-2)}(\nu-2+\beta(\nu-4) z_{L}^{2})=0.
\end{equation}
We find that like RS model there is no solution to satisfy the above
boundary conditions except for a trivial one, $(A,B)=(0,0)$
which is not of physical interest.\\
In this section we studied the VEV profile of a bulk scalar field under four boundary
conditions on a generalized warped brane-world model in the absence of brane localized potential.
We found that the VEV profiles depend on the vanishingly small brane cosmological
constant.\\

\section{ VEV profiles in the presence of brane localized potentials}
In this section, we formulate the VEV profiles of the bulk scalar
field in a case with brane localized scalar potentials in the
generalized Randall-Sundrum model. The action of a bulk field in the
presence of brane potentials reads as
\begin{equation}
S=\int d^{4}x \int_{0}^{L} dy
e^{-4A(y)}[-e^{2A(y)}|\partial_{\mu}\Phi|^{2}-|\partial_{y}\Phi|^{2}-V-\delta(y)V_{0}-\delta(y-L)V_{L}].
\end{equation}
The variation of the action is
$$\delta S=\int d^{4}x\int_{0}^{L} dy e^{-4A(y)}[\delta \Phi_{X}(\rho
\Phi_{X}-\frac{\partial V}{\partial \Phi_{X}})$$
\begin{equation}
+\delta (y)\delta \Phi_{X}(+ \partial_{y}\Phi_{X}-\frac{\partial
V_{0}}{\partial\Phi_{X}})+\delta (y-L)\delta \Phi_{X}(-
\partial_{y}\Phi_{X}-\frac{\partial
V_{L}}{\partial\Phi_{X}})].
\end{equation}
From above equation the Dirichlet boundary condition is the same as
$eq.(17)$ while the Neumann boundary condition should be modified as
\begin{equation}
\pm\partial_{y}\Phi_{X}-\frac{\partial
V_{\eta}}{\partial\Phi_{X}}|_{y=\eta=0,L}=0.
\end{equation}
In this paper we seek to generalize the Randall-Sundrum model and
moreover we want to investigate the GW mechanism. Therefore we use
the brane potentials as $[2,4,10]$
\begin{equation}
V_{\eta}=\lambda_{\eta}(|\Phi|^{2}-\frac{v_{\eta}^{2}}{2})^{2}=\frac{\lambda_{\eta}}{4}(\Phi_{R}^{2}+\Phi_{I}^{2}-v_{\eta}^{2})^{2},
\end{equation}
where $\lambda_{0}$ and $\lambda_{L}$ stand for the brane tension
and $\upsilon_{0}$ and $\upsilon_{L}$ are the VEV of the bulk scalar
field on the UV and IR brane respectively. By utilizing the above
scalar potentials we study the VEV profile in the case with brane
localized potential in the generalized RS model. We notice that like
the RS model, the $(D,D)$ boundary conditions in this case is
the same as one for $(D,D)$ case with brane potentials.\\
For some other choices of the brane potentials see $[8]$.

\subsection{$(D,N)$ case}
Here, we consider the $(D,N)$ case that can
 be given by
 \begin{equation}
\upsilon(1)=\upsilon_{1} \quad\quad\quad \partial_{z}
\upsilon(z)|_{z=z_{L}}+\frac{\partial
V_{L}}{\partial\Phi}|_{z=z_{L}}=0 .
\end{equation}
For AdS brane they are written down by
\begin{equation}
A(1+\alpha)+B(1+\beta)=\upsilon_{1},
\end{equation}
 and
$$k[Az_{L}^{\nu+2}(\nu+2+\alpha(\nu+4)
z_{L}^{2})-Bz_{L}^{-(\nu-2)}(\nu-2+\beta(\nu-4) z_{L}^{2})]+$$
\begin{equation}
\lambda_{L}(A z_{L}^{\nu+2}(1+\alpha z_{L}^{2})+B
z_{L}^{(-\nu+2)}(1+\beta z_{L}^{2}))[(A z_{L}^{\nu+2}(1+\alpha
z_{L}^{2})+B z_{L}^{(-\nu+2)}(1+\beta
z_{L}^{2}))^{2}-\upsilon_{L}^{2}]=0.
\end{equation}
We should remember that when the coupling in the boundary potential
is infinite, the Neumann boundary conditions turn to Dirichlet one
if we choose $\upsilon_{L}=\upsilon_{2}$.\\When the boundary quartic
coupling is finite, the numerical calculation indicates that $A\ll B
\simeq z_{L}^{\nu-2}\upsilon_{L}(1-\beta z_{L}^{2})$ as a solution
of $eq.(44)$ and $eq.(45)$. For the small boundary quartic coupling,
$A$ and $B$ can be approximated by
\begin{equation}
A \simeq
\chi B[\frac{\nu-2+\beta z_{L}^{2}(\nu-4)}{\nu+2+\alpha
z_{L}^{2}(\nu+4)}]z_{L}^{-2\nu}
\end{equation}
and
\begin{equation}
B\simeq\sqrt{\upsilon_{L}^{2}+\frac{k(1-\chi)[\nu-2+\beta
z_{L}^{2}(\nu-4)]}{\lambda_{L}(1+\beta z_{L}^{2})}}
\frac{z_{L}^{\nu-2}}{1+\beta z_{L}^{2}},
\end{equation}
where $\chi$ is determined by $eq.(44)$ and numerical calculation
shows that $\chi\simeq O(1)$ and $\chi=1$ for $\lambda_{L}=0$. Also,
$\chi$ is limited by
\begin{equation}
\chi\leq1+\frac{\lambda_{L}(1+\beta
z_{L}^{2})\upsilon_{L}^{2}}{k[\nu-2+\beta z_{L}^{2}(\nu-4)]}.
\end{equation}
\begin{figure}[htp]
 \begin{center}\includegraphics{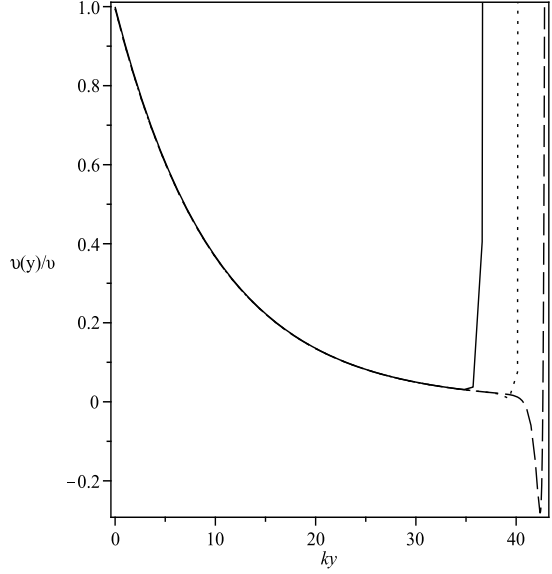}\vspace{6cm}
 \end{center}
  \caption{\small { The VEV profile for
 negative induced brane cosmological constant with
positive brane tension for $\aleph=33$ (dotted line), $\aleph=34$
(dashed line) and $\aleph=\infty$ (solid line i.e., RS case) in the
$(D,N)$ case for $\widetilde{\lambda}_{L}=100$.
 The values of other parameters of the model are the same as the ones in Figure $1$
.}}
 \end{figure}
In Fig. $3$ we show the dependence of VEV profile for negative
induced brane cosmological constant with positive brane tension with
$\widetilde{\lambda}_{L}=100$.\\
\subsection{$(N,D)$ case}
For the $(N,D)$ case, these boundary conditions given by
\begin{equation}
\partial_{z}
\upsilon(z)|_{z=1}-\frac{\partial V_{0}}{\partial\Phi}|_{z=1}=0
,\quad\quad\quad
\partial_{z} \upsilon(z)|_{z=z_{L}}=0.
\end{equation}
Then, for negative brane cosmological constant they are written down
by
$$k[A(\nu+2+\alpha (\nu+4))-B(\nu-2+\beta
(\nu-4))]$$
\begin{equation}
-\lambda_{0}(A(1+\alpha)+B(1+\beta))[(A(1+\alpha)+B(1+\beta))^{2}-\upsilon_{0}^{2}]=0,
\end{equation}
and
\begin{equation}
A z_{L}^{\nu+2}(1+\alpha z_{L}^{2})+B z_{L}^{(-\nu+2)}(1+\beta
z_{L}^{2})=\upsilon_{2}.
\end{equation}
Numerical calculation indicates that $A\sim
B\sim\frac{\upsilon_{2}}{z_{L}^{2}(1+\alpha z_{L}^{2})}$ and for
small $\lambda_{0}$, $\widetilde{\lambda}_{0}\leq O(0.01)$, $A$ and
$B$ are approximated by
\begin{equation}
A\simeq(1-\frac{\Gamma_{1}}{\Gamma_{1}+\Gamma_{2}
z_{L}^{2\nu}})\frac{\upsilon _{2} z_{L}^{-(\nu+2)}}{1+\alpha z_{L}^{2}},
\end{equation}
and
\begin{equation}
B\simeq(\frac{\Gamma_{1}}{\Gamma_{1}+\Gamma_{2}
z_{L}^{2\nu}})\frac{\upsilon _{2} z_{L}^{\nu-2}}{1+\beta z_{L}^{2}}.
\end{equation}
Where
\begin{equation}
\Gamma_{1}=(1+\beta
z_{L}^{2})[k(\nu+2+\alpha(\nu+4))+\lambda_{0}\upsilon_{0}^{2}(1+\alpha)],
\end{equation}
and
\begin{equation}
\Gamma_{2}=(1+\alpha
z_{L}^{2})[k(\nu-2+\beta(\nu-4))-\lambda_{0}\upsilon_{0}^{2}(1+\beta)].
\end{equation}
\begin{figure}[htp]
 \begin{center}\includegraphics{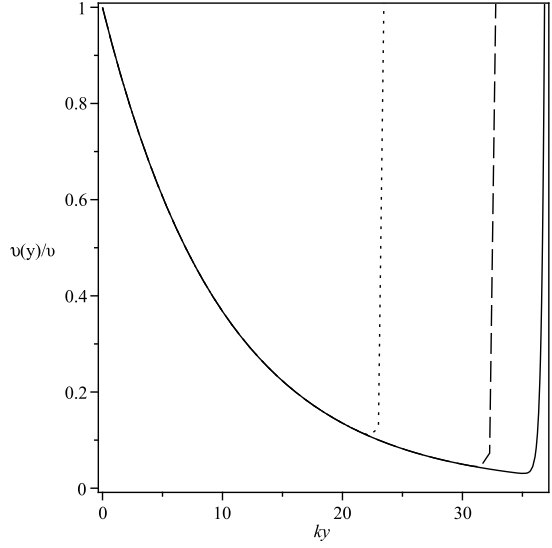}\vspace{6cm}
 \end{center}
  \caption{\small { The VEV profile for large
positive brane cosmological constant with $\aleph=20$ (dotted line),
$\aleph=28$ (dashed line) and $\aleph=\infty$ (solid line i.e., RS
case)in the
$(N,D)$ case for $\widetilde{\lambda}_{0}=100$.
 The values of other parameters of the model are the same as the ones in Figure $1$
.}}
 \end{figure}
 Figure $4$ shows the
VEV profile for large positive brane cosmological constant,
 when
$\widetilde{\lambda}_{0}=100$.\\
\subsection{$(N,N)$ case}
The $(N,N)$ case of boundary conditions for AdS brane are
\begin{equation}
\partial_{z}
\upsilon(z)|_{z=1}-\frac{\partial V_{0}}{\partial\Phi}|_{z=1}=0
,\quad\quad\quad
\partial_{z}
\upsilon(z)|_{z=L}+\frac{\partial V_{L}}{\partial\Phi}|_{z=L}=0.
\end{equation}
Then, the boundary conditions can be written down as
$$k[A(\nu+2+\alpha (\nu+4))-B(\nu-2+\beta
(\nu-4)]$$
\begin{equation}
-\lambda_{0}(A(1+\alpha)+B(1+\beta))[(A(1+\alpha)+B(1+\beta))^{2}-\upsilon_{0}^{2}]=0,
\end{equation}
and
$$k[Az_{L}^{\nu+2}(\nu+2+\alpha(\nu+4)
z_{L}^{2})-Bz_{L}^{-(\nu-2)}(\nu-2+\beta(\nu-4) z_{L}^{2})]+$$
\begin{equation}
\lambda_{L}(A z_{L}^{\nu+2}(1+\alpha z_{L}^{2})+B
z_{L}^{(-\nu+2)}(1+\beta z_{L}^{2}))[(A z_{L}^{\nu+2}(1+\alpha
z_{L}^{2})+B z_{L}^{(-\nu+2)}(1+\beta
z_{L}^{2}))^{2}-\upsilon_{L}^{2}]=0.
\end{equation}
\begin{figure}[htp]
 \begin{center}\includegraphics{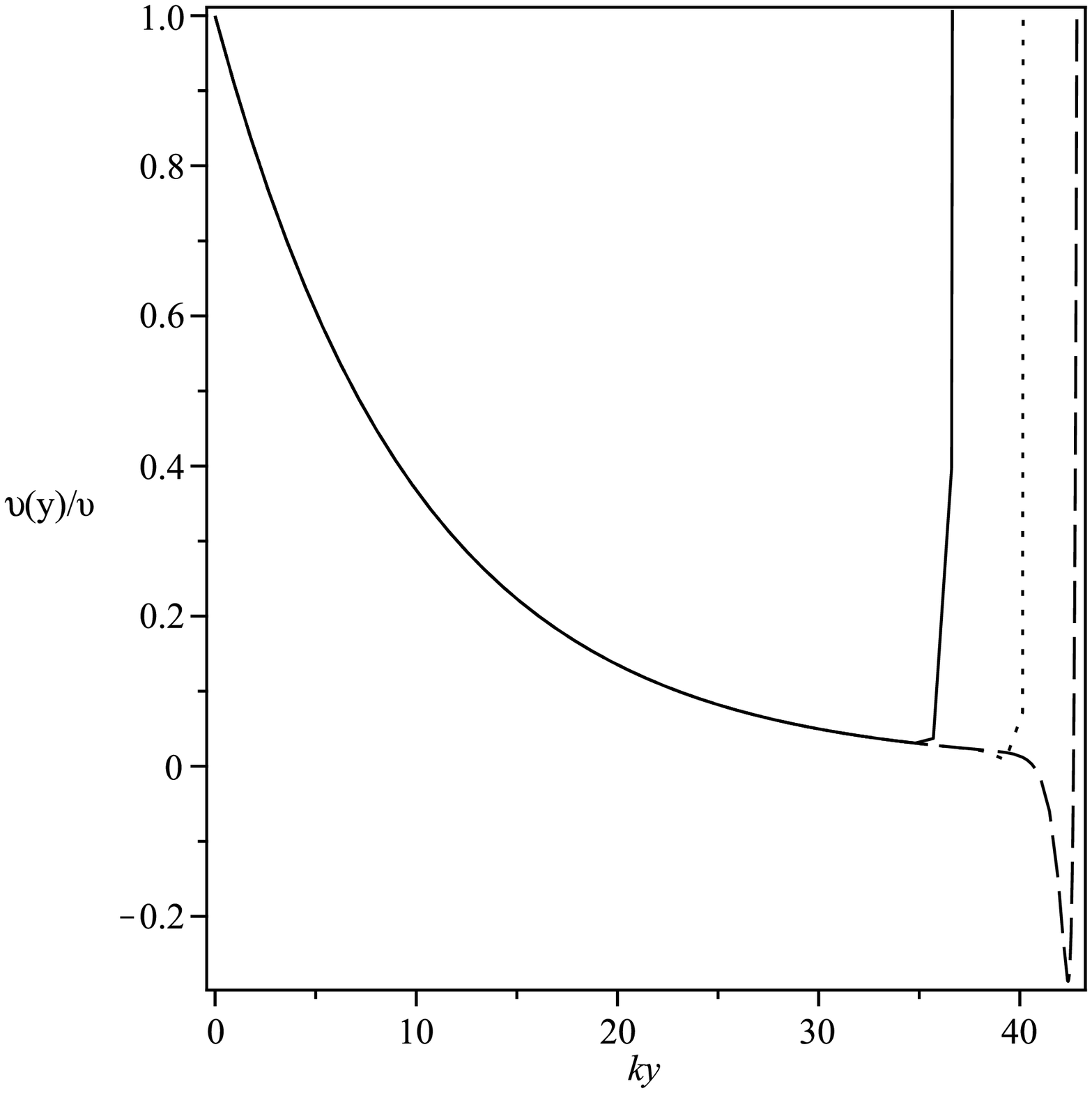}\vspace{6cm}
 \end{center}
  \caption{\small {The VEV profile
for negative induced brane cosmological constant with
positive brane tension for $\aleph=33$ (dotted line), $\aleph=34$
(dashed line) and $\aleph=\infty$ (solid line i.e., RS case) in the
$(N,N)$ case for
$\widetilde{\lambda}_{0}=100,\widetilde{\lambda}_{L}=100$.
 The values of other parameters of the model are the same as the ones in Figure $1$
.}}
 \end{figure}
In the limit of very small boundary couplings,
$\widetilde{\lambda}_{0}\ll O(0.01)$ and $\widetilde{\lambda}_{L}\ll
O(0.1)$, $A$ and $B$ are zero but for large boundary quartic
couplings which is considered in GW mechanism we have
\begin{equation}
A_{GW}\simeq -\upsilon_{0}\frac{(1+\beta
z_{L}^{2})}{\Delta_{1}}z_{L}^{-2\nu}+\upsilon_{L}\frac{(1+\beta)}{\Delta_{1}}z_{L}^{-(\nu+2)},
\end{equation}
and
\begin{equation}
B_{GW}\simeq \upsilon_{0}\frac{(1+\alpha
z_{L}^{2})}{\Delta_{1}}(1+\frac{\Delta_{2}}{\Delta_{1}}z_{L}^{-2\nu})-\upsilon_{L}\frac{(1+\alpha)}{\Delta_{1}}z_{L}^{-(\nu+2)}.
\end{equation}
Where
\begin{equation}
\Delta_{1}=(1+\beta)(1+\alpha z_{L}^{2}),
\end{equation}
and
\begin{equation}
\Delta_{2}=(1+\alpha)(1+\beta z_{L}^{2}).
\end{equation}
Here we numerically analyze the VEV profile for finite boundary
couplings.
 The result is shown in Figure $5$ for large values
of $\widetilde{\lambda}_{0}$ and $\widetilde{\lambda}_{L}$.
At the end of this section we reconsider the realizations of the GW
mechanism under general boundary conditions in the generalized RS
model. In the GW mechanism the effective potential for the modulus
can be generated by the total action integral of a bulk scalar field
with quartic interactions localized in two $3-$branes. The bulk
scalar field acts an important role to stabilize the radion which
yields a compactification scale in terms of VEVs profile of the
scalar field
at two branes. The VEV of the bulk scalar can be obtained by solving the equation of motion.\\
Recently, this mechanism was studied in the generalized warped brane
models with a nonzero brane cosmological constant by $[9]$. They
have been obtained the modulus stabilization condition both for
positive
and negative values of the brane cosmological constant which increases (decreases) from RS value with increasing $\omega^{2}$.\\
we have investigated the unknown coefficients A and B of the VEV
profile of the bulk scalar field in the generalized RS model by
imposing boundary conditions which are given in the sections $3$ and
$4$. It is well known that the $(N,N)$ type boundary conditions
given by $eq.(59)$ and $(60)$ should be taken in the GW mechanism.
The situation in the generalized warped model is similar to RS
warped model as discussed in Ref. $[10]$. When the boundary quartic
couplings $\lambda_{0}$ and $\lambda_{L}$ are large the $(N,N)$ BCs
which imposed in the GW mechanism are equivalent to $(D,D)$ ones
given in $(20)$ if we take $\upsilon_{1}=\upsilon_{0}$ and
$\upsilon_{2}=\upsilon_{L}$, so by this assumption in the limit of
large boundary couplings, A and B given by $eq.(21)$ and $eq.(22)$
are $A\simeq A_{GW}$ and $B\simeq B_{GW}$, therefore the GW
mechanism can be realized in the $(D,D)$ case with appropriate
VEV profile.\\
In the generalized warped model as mentioned in Ref. $[10]$, when
the brane localized potential is considered, GW mechanism can also
work in $(N,D)$ and $(D,N)$ BCs with large boundary quartic
couplings. The GW mechanism can not work when one of the boundary
coupling $\lambda_{0}$ or $\lambda_{L}$ becomes small in $(N,N)$
case.

\section{Conclusions}
In the RS model a negative tension visible brane is utilized to
describe our Universe. However, Such negative tension branes are
unstable. Hence in this work we considered a generalized RS warped
model.\\
 In this paper we investigated a bulk scalar field under Dirichlet
and Neumann boundary conditions in a generalized RS warped model
with a non zero cosmological constant on the brane. First we have
obtained the profiles of VEV of the bulk scalar in the absence of
brane localized potential for the negative and positive brane
cosmological constant. We find that: \\
$(1)$ For an AdS visible brane with negative brane tension the
results
are similar to RS case.\\
$(2)$ However for an AdS visible brane with positive brane tension
the results are different from the RS case. In this case the VEV
profile undergo sudden changes near the IR brane. Moreover we see
that the IR brane is displaced from it's RS location. The
coefficients $A$ and $B$ are functions of the length of the
stabilized modulus. Hence the change of the length of the stabilized
modulus modifies the shape of the VEV in comparison to the RS case.
\\
$(3)$ For a dS visible brane the results for small values of
$\omega^2$ are not different from the RS case. But for larger values
of $\omega^2$ we have departures from the RS result.\\
Next we have considered the VEV profiles under general boundary
conditions in the presence of brane localized potential for non zero
brane cosmological constant. We find that to see drastic changes of
the profile near the IR brane for an AdS visible brane with positive
brane tension
 the
value of $\widetilde{\lambda}_{L}\sim O(1)$ in the RS case but the
value of $\widetilde{\lambda}_{L}\sim O(10)$ in the present case.\\
 We have also verified GW mechanism in the generalized RS model
with Dirichlet and Neumann Boundary conditions. Analogous to RS
model $[10]$ the GW mechanism can work under non zero Dirichlet BCs
with appropriate VEVs in the generalized RS model. It will be
interesting to use these results to study the Dirichlet Higgs as
radion stabilizer in the generalized
warped campactification.\\
Our analysis of consistent boundary conditions is similar to that of
Csaki, et. al. $[17]$. It is possible to extend this work by
studying
 the behavior of the bulk scalar field on the warped five
dimension by utilizing the background field method, separating the
field into classical and quantum fluctuation parts and to
investigate
the phenomenological consequences of the profile of the scalar field $[10]$. \\
There are quantum corrections, from Casimir effect and induced
gravity on the brane. One requires that the brane world metric be a
solution of the quantum-corrected Einstein equations. For the RS
model this issue has been addressed in Refs. $[18,19]$. From
$eq.(12)$ we see that in the generalized RS model $A(y)=k|y|$ plus
extra small term due to the curved branes. In the RS model the sum
of brane tensions of two branes are zero. However in the generalized
RS model this sum is not zero $[3]$. Hence the discussion of the
issue of self-consistency is different from that of the RS model.
However it is reasonable to expect that GRS be self consistent for
some specific values of the parameters of the model
.\\
 The
analysis presented in this work we have ignored the backreaction of
the scalar field on the spacetime geometry. The stress tensor for
the scalar field in the limit of large modulus can be obtained $[4]$
. For the $(N,N)$ case with large boundary coupling,stress tensor
for the scalar field can we find that if $v^2_0\Delta^2_1/M_5^3<<1$,
$v^2_L\Delta^2_1/M_5^3<<1$ and $m^2/k^2<<1$, then one can neglect
the stress tensor for the scalar field in comparison to the stress
tensor induced by the bulk cosmological constant. Furtheremore the
parameters of the bulk and brane potential are constrained by these
inequalities.
\\
In this work our choices for the bulk and the brane potentials were
those of Ref. $[2]$. Other choices for the bulk potential and brane
potentials are possible. The potential for the $GW$ scalar $\Phi$ in
the bulk $[20]$ has the general form of
$V(\Phi)=m^2\Phi^2+C_3\Phi^3+C_4\Phi^4+......$ . It would be
interesting to study the VEV profiles of the bulk scalar field in
such generalized warped
models.\\
In a previous work we studied the case of a scalar field
non-minimally coupled to the Ricci curvature scalar $[21]$. The case
of massless, conformally coupled scalar field are self-conistent
 after the inclusion
of quantum corrections. Hence it will be of interest to consider
 massless, conformally coupled scalar field in the generalized RS
 framework.\\\\
\clearpage
{\bf{Appendix : The variation of $kr\pi$ and the brane tensions}}\\\\
In the $RS$ $[2]$ case the brane tension of the visible brane is
negative
and to solve the hierarchy problem the value of $kr\pi=36.84$.\\
The situation is different in the generalized $Randall-Sundrum$
model.\\
This subject is discussed in Ref. $[3]$. Here we provide a concise
 presentation.\\

 {\bf The AdS case}\\\\
If we denote $\varsigma=kr\pi$ then for the case where the
cosmological constant of the brane is negative we have $[3]$
\begin{equation}
e^{-\varsigma}=\frac{10^{-n}}{c_1}[1\pm\sqrt{1+\omega^210^{2n}}] ,
\end{equation}
with $c_{1}=1+\sqrt{1-\omega^{2}}$. Let $\varsigma_1$ and
$\varsigma_2$ corresponds to the plus and minus sign respectively.
The $\varsigma_1$ case is not much different from the $RS$ case. But
the $\varsigma_2$ is quite distinct from the $RS$ case. In the limit
$\aleph-2n\gg 1$
\begin{equation}
\varsigma_1\simeq nln10+\frac{1}{4}10^{-(\aleph-2n)} \qquad
\varsigma_2\simeq (\aleph-n)ln10+ln4 ,
\end{equation}
the value of $\varsigma_2$ can
be as large as $250$.\\
 Next we will show below in this
situation the brane tension of the visible brane is positive. As far
as the stability of the brane is concern this is a desirable
feature.
 To compute the brane tension $U_{vis}$ of the AdS case
 we rewrite the brane tension of the visible brane of Ref. $[3]$ as
\begin{equation}
U_{vis}=12M^3k[\frac{1-(c_1e^{-\xi}\omega^{-1})^2}{1+(c_1e^{-\xi}\omega^{-1})^2}].
\end{equation}
Now inserting $eq.(63)$ in $eq.(65)$ and with
$\omega^2=10^{-\aleph}$, the correct form of brane tension of the
visible brane is
\begin{equation}
U_{vis}=(12M^3k)\frac{1-10^{\aleph-2n}(1\pm\sqrt{1-10^{-(\aleph-2n)}})^2}{1+10^{\aleph-2n}(1\pm\sqrt{1-10^{-(\aleph-2n)}})^2}.
\end{equation}
And finally from this result we can compute the brane tension in the
limit where $\aleph-2n\gg 1$. For the plus sign case the brane
tension for the visible brane is negative and it is given
\begin{equation}
U_{vis-1}\simeq -12M^3k.
\end{equation}
Which is in agreement with the result of Ref. $[3]$.\\
But for the minus sign case the brane tension for the visible brane
is positive and it is given
\begin{equation}
U_{vis-2}\simeq 12M^3k.
\end{equation}
Which differs from the result of Ref. $[3]$ by a factor $3$.\\

{\bf The dS case}\\\\
When the cosmological constant of the brane is positive we have
\begin{equation}
e^{-\varsigma}=\frac{10^{-n}}{c_2}[1+\sqrt{1+\omega^210^{2n}}] ,
\end{equation}
with $c_{2}=1+\sqrt{1+\omega^{2}}$. It turns out that the brane
tension of the visible brane is negative for the entire range of the
positive values of $\Omega$. However the difference with the $RS$
case is due to the value of $\varsigma$.  For instance for
$\aleph=20$ we obtain $\varsigma=23.7$.\\

\end{document}